\documentclass[letterpaper, 10 pt, conference]{ieeeconf}  

\IEEEoverridecommandlockouts                              

\usepackage{graphics} 
\usepackage{epsfig} 
\usepackage{mathptmx} 
\usepackage{times} 
\usepackage{amsmath} 
\usepackage{amssymb}  
\usepackage{xcolor}
\definecolor{ours}{HTML}{1F77B4}
\definecolor{NH}{HTML}{FF7F0E}
\definecolor{NC}{HTML}{2CA02C}
\definecolor{NA}{HTML}{7837b3}
\definecolor{RA}{HTML}{D62728}
\definecolor{BC}{HTML}{7F7F7F}

\usepackage{amsfonts}
\usepackage{tabularx}
\usepackage{bm}
\usepackage{mathtools}
\usepackage[ruled, vlined, linesnumbered]{algorithm2e}
\usepackage[utf8]{inputenc}
\usepackage{cite}
\usepackage{authblk}
\providecommand{\shortcite}[1]{\cite{#1}}

\newtheorem{definition}{Definition}[section]

\title{\LARGE \bf
Resource-Aware Adaptation of \\ Heterogeneous Strategies for Coalition Formation
}

\author{Anusha Srikanthan$^{\dagger}$, Harish Ravichandar$^{\dagger}$
\thanks{*This work was supported by the Army Research Lab under Grant W911NF-20-2-0036.}
\thanks{$^{\dagger}$Georgia Institute of Technology {\tt\small \{asrikanthan3, harish.ravichandar\}@gatech.edu}}%
}

\begin{document}

\maketitle
\thispagestyle{empty}
\pagestyle{empty}

\begin{abstract}

Existing approaches to coalition formation often assume that requirements associated with tasks are precisely specified by the human operator. However, prior work has demonstrated that humans, while extremely adept at solving complex problems, struggle to \textit{explicitly} state their solution strategy. Further, existing approaches often ignore the fact that experts may utilize different, but equally-valid, solutions (i.e., \textit{heterogeneous strategies}) to the same problem. In this work, we propose a two-part framework to address these challenges. First, we tackle the challenge of inferring \textit{implicit} strategies directly from expert demonstrations of coalition formation. To this end, we model and infer such heterogeneous strategies as capability-based requirements associated with each task. Next, we propose a method capable of \textit{adaptively} selecting one of the inferred strategies that best suits the target team without requiring additional training. Specifically, we formulate and solve a constrained optimization problem that \textit{simultaneously} selects the most appropriate strategy given the target team's capabilities, and allocates its constituents into appropriate coalitions. We evaluate our approach against several baselines, including some that resemble existing approaches, using detailed numerical simulations, StarCraft II battles, and a multi-robot emergency-response scenario. Our results indicate that our framework consistently outperforms all baselines in terms of requirement satisfaction, resource utilization, and task success rates.

\end{abstract}

\section{INTRODUCTION}

In this work, we focus on the \textit{coalition formation} problem in which a team of robots need to be divided or partitioned into non-overlapping sub-teams (i.e., coalitions) such that multiple concurrent tasks can be successfully carried out~\cite{gerkey2004formal,korsah2013comprehensive}. In particular, we are interested in coalition formation for \textit{heterogeneous teams} (i.e., teams made of robots with different capabilities). Forming effective coalitions with heterogeneous multi-robot systems has been proven to be effective in several domains, such as environmental monitoring~\cite{shkurti2012multi}, agriculture~\cite{tokekar2016sensor}, and construction~\cite{werfel2014designing}.  

Most existing approaches to coalition formation assume that the task requirements associated with different tasks are explicitly specified by the human operator (e.g.,~\cite{prorok2017impact,ravichandar2020strata}). However, prior work has demonstrated that manually specifying multi-dimensional objective functions that capture the trade-offs between multiple factors can be very challenging~\cite{nisbett1977telling, rieskamp2003people, rieskamp2006ssl}. In fact, complex tasks involve multi-dimensional requirements (e.g., searching an area, carrying a certain weight, and navigating rugged terrain) that are hard to explicitly and precisely specify.

Despite the challenges in explicit specification of requirements and objectives, domain experts routinely solve complex problems with similar requirements~\cite{nisbett1977telling}. For example, the field of Inverse Reinforcement Learning (IRL), which is based on the finding that users have an easier time \textit{demonstrating} a task than writing a reward function, has led to successful applications in autonomous flight~\cite{abbeel2004apprenticeship}, social navigation~\cite{kretzschmar2016socially}, and manipulation~\cite{kalakrishnan2013learning}. 

\begin{figure}[t]
    \centering
    \includegraphics[width=\columnwidth]{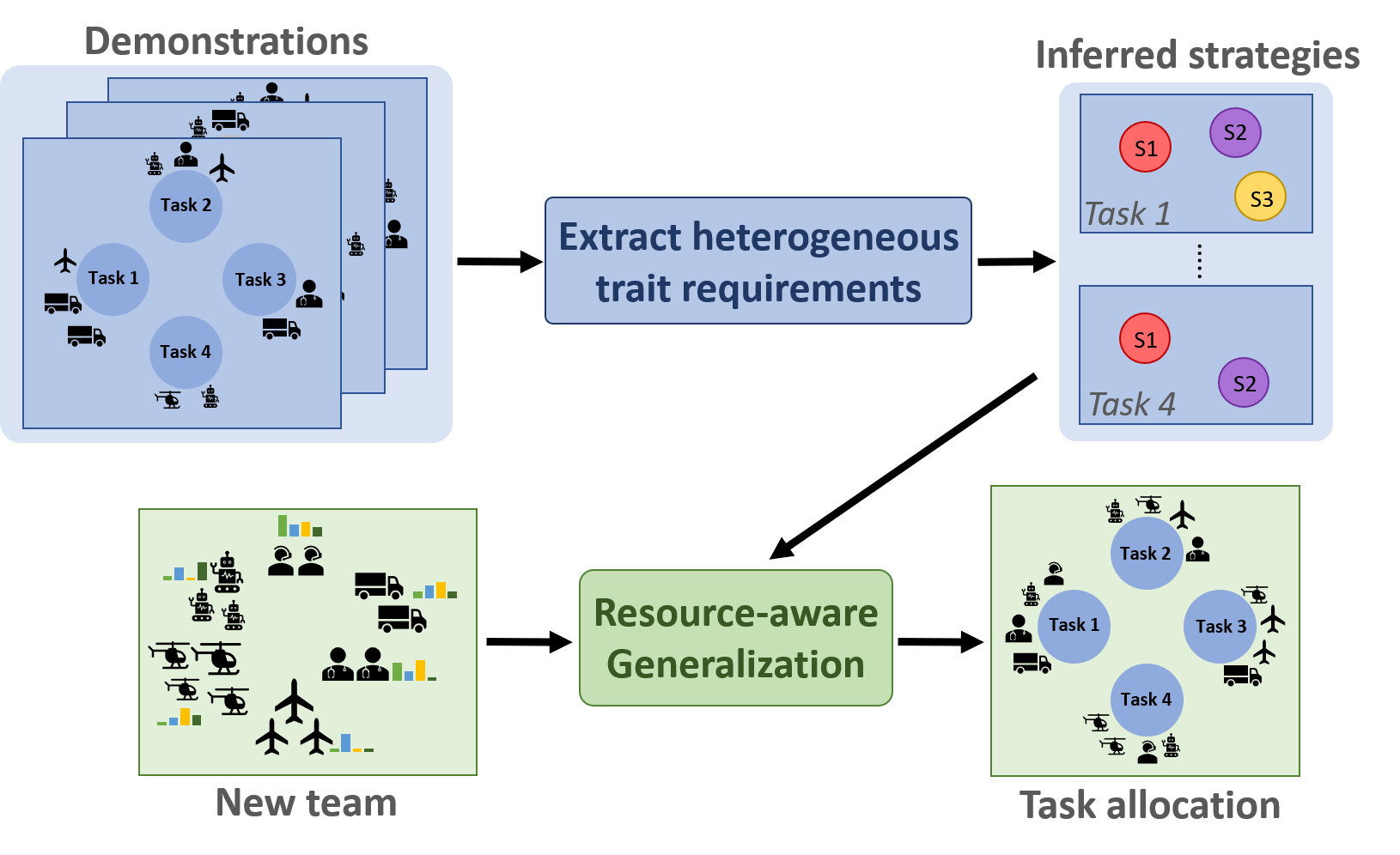}
    \caption{The block diagram represents our approach of learning implicit heterogeneous strategies from expert demonstrations of coalition-task pairs. On top, the figure shows our hierarchical clustering-based method to compute heterogeneous requirements from expert demonstrations. The bottom half of the figure is our generalization framework based on the optimization of coalitions paying explicit attention to the compatibility of the team's capabilities with the inferred strategies for each task.}
    \label{fig:block_diagram}
\end{figure}

In this work, we take an IRL-inspired approach to coalition formation. Instead of requiring exact task requirements, we propose to leverage \textit{expert demonstrations} of coalition formation. Our approach is similar in spirit to IRL in that we attempt to infer the task requirements that the expert was implicitly attempting to satisfy when providing the demonstrations. Specifically, we define task requirements in terms of the \textit{traits} (i.e., multi-dimensional capabilities) required to accomplish the task. We then formulate our problem as determining i) the \textit{trait requirements} of each task from expert demonstrations, and ii) the \textit{allocation} of robots to each task such that each coalition collectively satisfies the inferred requirements.

A key challenge in learning from human demonstrations is that different users may demonstrate different, equally-valid, solutions to a task. 
A rich body of work in psychology and human-robot teaming suggests that complex tasks are often solved using one of many comparable strategies~\cite{payne1992constructive,reverdy2014modeling}. For instance, one can search an area either with i) a slow-moving coalition with a large collective sensing area, or with ii) a fast-moving coalition with a small collective sensing area. We denote such different, yet equivalent, trait requirements as \textit{heterogeneous strategies}. To understand why prior work fails to efficiently handle the existence of \textit{heterogeneous strategies}, let us assume that all the strategies are known beforehand and extend the approach contributed in \cite{ravichandar2020strata} to optimize for the coalition-task pairs. In such a scenario, there would be a combinatorially large space of possible strategies to choose from for each task with no additional information on how to select the appropriate strategies. It would be computationally inefficient to use brute force search to collectively identify the set of strategies for all tasks that achieve the best performance. 

Therefore, in this work, we contribute: 1) a novel formulation of heterogeneous coalition formation strategies based on trait requirements, 2) a clustering-based approach for inferring such generalizable heterogeneous strategies from expert demonstrations, and 3) an optimization-based method to resource-aware selection of strategies to generalize the inferred strategies to entirely new target teams without additional training.

We propose a hierarchical clustering-based method to extract heterogeneous requirements from expert demonstrations consisting of successful assignments (see top half of Fig. \ref{fig:block_diagram}). The clustering approach provides appropriate \textit{abstraction} of trait requirements by distilling expert demonstrations into a small number of strategies. Additionally, clustering the expert demonstrations separately for each task filters the noise in the individual demonstrations leading to better generalization which we aim to show through our experiments by considering a baseline without abstraction.

To enable \textit{resource-aware} generalizations of the inferred strategies to new target teams, we  pay explicit attention to the compatibility of the team's capabilities with the inferred strategies for each task (see bottom half of Fig. \ref{fig:block_diagram}). We propose an optimization-based approach that \textit{simultaneously} i) selects the strategy for each task and ii) optimizes the constituents of the coalitions, both based on the \textit{context} of the target team's capabilities and trait requirements. 

Our framework accounts for task-interdependence by solving a constrained optimization problem to \textit{simultaneously} optimize both strategy selection and robot assignment. This is important due to the fact that the team has finite resources, and as such, the selection of strategy for one task affects the selection for other tasks.

We validate our contributions by evaluating our approach against a variety of baselines~\cite{ravichandar2020strata,vig2006multi, duvallet2010imitation,cui2017learning}, using detailed numerical experiments, battles in StarCraft II game, and a disaster response scenario on the Robotarium~\cite{wilson2020robotarium}. Our results demonstrate that our approach consistently outperforms the baseline approaches in terms of requirement satisfaction, resource utilization, and task success rates.


\section{RELATED WORK}

We discuss prior and related work from the following broad categories to contextualize our contributions.

\subsection{Task allocation} 

Formally, task allocation problems are categorized based on three axes: Single-Task (ST) vs. Multi-Task (MT) robots; Single-Robot (SR) vs. Multi-Robot (MR) tasks; and Instantaneous Allocation (IA) vs. Time-extended Allocation (TA). We refer readers to Korsah et al. \shortcite{korsah2013comprehensive} for a comprehensive categorization survey. Our work addresses the coalition formation problem, which is an instance of the ST-MR-IA problem. 

We note that the coalition formation problem is closely related to the problems of multi-agent planning~\cite{torreno2017cooperative} and scheduling~\cite{gombolay2016apprenticeship}. However, each of these problems differ considerably in terms of their objectives and what is assumed to be given. Specifically, planning approaches assume access to domain definitions and action models, while scheduling approaches assume access to the known set of actions to take. In contrast, coalition formation approaches do not assume access to such information and focus on optimizing the constituents of each coalition such that task requirements are satisfied. As such, we view our work as complementary to existing approaches for planning and scheduling. 

\subsection{Coalition formation} 

Existing approaches to coalition formation fall into one of three groups: utility-based, auction-based, and trait-based methods. First, prior work has demonstrated that utility-based methods can compute optimal coalitions based on the coalition values for each team assigned to a task~\cite{shehory1995task,vig2006multi,afghah2018coalition,zhang2013considering}. However, the design of utility functions requires considerable domain knowledge and often needs to be hand-crafted for the particular set of tasks and robots. Second, auction-based approaches have had tremendous success in solving a wide range of coalition formation problems using mechanisms through which robots bid on tasks~\cite{guerrero2003multi,lin2005combinatorial,xie2018mutual}. A common limitation of auction methods is that they rely on extensive communication for bidding and scale poorly with the number of robots in the team. Lastly, recent advances in task allocation have introduced methods that model both robots and tasks in terms of their traits (i.e., capabilities)~\cite{prorok2017impact,ravichandar2020strata, mayya2020resilient}. We adopt a similar approach and model our robots and tasks in terms of traits. A unique benefit of using trait-based approaches is that they are generalizable to new teams, as task requirements are specified in terms of traits, and not specific robots.

We deviate from prior work in coalition formation in terms of two crucial assumptions. First, methods discussed thus far assume complete knowledge of task requirements, whatever the form. Second, most assume that there is exactly one set of task requirements (i.e., no heterogeneity). Authors of \cite{mayya2020resilient} propose a method for resilient task allocation by considering more than one set of task requirements, however, assume that such requirements are known. 
In contrast, we introduce a model for heterogeneous task requirements and contribute an algorithm that extracts such implicit requirements directly from expert demonstrations. Further, our approach satisfies task requirements by jointly optimizing strategy selection and robot assignment. 

\subsection{Heterogeneous learning from demonstrations}

Next, we examine related work focused on learning from heterogeneous demonstrations. Recent advances in learning from demonstrations has introduced methods to capture heterogeneity from expert demonstrations. Prior work has showcased the ability to infer discrete modes of operation~\cite{li2017infogail}, multiple visual intentions~\cite{tamar2018imitation}, and most commonly diverse user preferences~\cite{dimitrakakis2011bayesian,nikolaidis2015efficient,paleja2019heterogeneous,chen2020joint} from heterogeneous demonstrations. Existing approaches that embrace heterogeneity in demonstrations focus on inferring and encoding the distinct characteristics. In contrast, our approach focuses on optimizing the selection of the most appropriate inferred strategy based on the available resources. Further, our work represents the first attempt to learn heterogeneous task allocation strategies from demonstrations.

\subsection{Learning for team-level coordination} 

Finally, we discuss attempts in the general area of multi-agent learning~\cite{panait2005cooperative} and specifically scrutinize the details of learning methods for task allocation. Much of the literature on learning for multi-agent systems is focused on reinforcement learning applied to low-level coordination (e.g.,~\cite{gupta2017cooperative,lowe2017multi}). However, only a handful of methods exist for learning high-level coordination from demonstrations. Here, we discuss a few notable examples. An approach introduced in~\cite{duvallet2010imitation} models task features and learns their weights from expert demonstrations to bias the prices in a market-based task allocation algorithm. A recent structured prediction approach \cite{carion2019structured} for task allocation uses a combination of reinforcement learning and quadratic integer programming for learning directly from data to optimize assignments. The approaches in~\cite{duvallet2010imitation,carion2019structured}, however, assume the existence of a single strategy for task allocation. A distributed approach for multi-agent task allocation~\cite{turner2018distributed} learns to select the most appropriate of two pre-specified strategies, namely Earliest Deadline First (EDF) or Nearest Task First (NTF). Finally, reinforcement learning has been utilized to address repeated coalition formation problems that involve teammates whose capabilities are initially unknown~\cite{chalkiadakis2012sequentially}.

The learning methods discussed so far consider only homogeneous robots~\cite{carion2019structured}, do not show generalization to teams unseen during training~\cite{duvallet2010imitation,cui2017learning}, depend on the ability to interact with the environment to learn policies~\cite{chalkiadakis2012sequentially} or adhere to a limited number of pre-specified strategies~\cite{turner2018distributed}. In contrast, our framework is capable of learning generalizable and heterogeneous strategies for task allocation in heterogeneous multi-agent systems from expert demonstrations, and do not rely on environmental interactions.


\section{PROBLEM FORMULATION}

We consider a heterogeneous team composed on $S$ \textit{species} (i.e., robot types), in which the $s^{\text{th}}$ species contains $N_s$ robots. Let the team be tasked with a set of $M$ concurrent tasks denoted by $\mathcal{T} = \{T_1, T_2, .., T_M\}$. We now introduce a series of pertinent definitions.

\begin{definition}\label{def:species}
The \textbf{capabilities} of the team are encoded by its Species-Trait matrix $Q = [q_1, \cdots q_S]^T \in \mathbb{R}_+^{S \times U}$, where $q_s \in \mathbb{R}^U_+$ is the trait vector listing the different traits of the $s^{\text{th}}$ species, and $U$ is the total number of traits.
\end{definition}

\begin{definition}\label{def:assignment}
The \textbf{assignment} of robots to tasks is encoded by the Assignment matrix $X =[x_{ms}] \in \mathbb{Z}_+^{MxS}$, where the $ms^{\text{th}}$ element $x_{ms}$ denotes the number of robots from Species $s$ assigned to Task $T_m$.
\end{definition}

\begin{definition}\label{def:aggregation}
The \textbf{aggregated traits} at each task  due to the assignment of robots is encoded in the Task-Trait matrix $Y \in \mathbb{R}_+^{M \times U}$, whose $m^{\text{th}}$ row represents the aggregation of traits for Task $T_m$ and is given by
\begin{equation}
    y_{m} = Q^Tx_{m},\ \forall m \in \{1, \cdots, M\}
\end{equation}
\end{definition}

\begin{definition}\label{def:strategies}
Let a set of heterogeneous \textbf{strategies} associated with the $m^{\text{th}}$ task be made up of $P_m$ strategies. Each strategy denotes the trait requirements, when satisfied, leading to successful completion of the task. As such, the $r^{\text{th}}$ strategy for the $m^{\text{th}}$ task is given by
\begin{equation}\label{eq:strategies}
    \prescript{r}{}{y}^*_m \in \mathbb{R}_+^U, \forall r \in \{1,\cdots,P_m\} 
\end{equation}
where $P_m$ represents the number of strategies for Task $T_m$.
\end{definition}

Note that our definition of strategies allows for different sets of requirements to be associated with the same task. We assume that satisfying any one of the $P_m$ task requirements will result in successful completion of the $m^{\text{th}}$ task.

\begin{definition}\label{def:demonstrations}
Let a set of $N$ expert \textbf{demonstrations} in the form of robot assignments be given by
\begin{equation}\label{eq:dataset}
    \mathcal{D} = \{X^{(i)},\ Q^{(i)}\}_{i=1}^N
\end{equation} 
where $X^{(i)}$ represents the expert-specified assignment matrix for a team with capabilities encoded by the Species-Trait matrix $Q^{(i)}$. 
\end{definition}

Given the above definitions, our problem consists of two steps: \textit{i)} extracting the set of approximated strategies $^r \hat{y}_m$ for each task from $\mathcal{D}$, and \textit{ii)} optimizing the Assignment matrix $X^{(j)}$ of a new team with Species-Trait matrix $Q^{(j)} \notin \mathcal{D}$ such that the associated $y_m \succ\ ^r \hat{y}_m,\ \forall m$, where $y_m$ denotes the traits aggregated by the target team towards the $m^{\text{th}}$ task and $\succ$ denotes the element-wise greater-than operator.


\section{MULTI-STRATEGY COALITION FORMATION}

In this section, we provide details of our proposed framework. We peruse a running example to illustrate each component of our framework. To this end, let us consider a disaster response mission involving tasks such as removing debris, searching urban environments, and retrieving objects or people from damaged buildings. 

\subsection{Extracting heterogeneous task requirements}\label{subsec:extraction}
In order to extract implicit task requirements, we first compute the trait aggregations from the demonstrations $\mathcal{D}$ as
\begin{equation}
    y_m^{(i)} = {Q^{(i)}}^T x_m^{(i)}, \quad \forall m, i \label{eq:demo_trait_agg}
\end{equation}
where $y_m^{(i)}$ represents trait aggregation associated with the $m^{\text{th}}$ task of the $i^{\text{th}}$ demonstration. Given that there might be multiple strategies to solve each task, the computed trait aggregations in (\ref{eq:demo_trait_agg}) are likely to form distinct clusters in the trait space, with each cluster representing a unique set of trait requirements or strategy.

Note that we do not attempt to capture the specific robots (e.g., 5 quadrupeds and 3 aerial vehicles) used by an expert. Instead, we extract the amounts of specific traits that are possessed by the experts' coalitions (e.g., 1 $km^2$ aggregate coverage area, and 5 robots with ability to navigate rugged terrain). Adopting these trait-based specifications allows us to generalize the extracted requirements to new teams containing entirely new kinds of robots.

We then apply agglomerative (i.e., hierarchical bottom-up) clustering on the computed trait aggregation vectors for each task to obtain clusters denoted by $\{\prescript{r}{}{C_m}\}_{r=1}^{P_m}$. While our framework is agnostic to the specific clustering approach, hierarchical clustering allows for adaptively choosing the number of clusters based on the task's characteristics. On a related note, the choice of the neighborhood distance thresholds can be made from analyzing the dendrogram plot for each task. 

Once the clusters are identified, we compute the approximate task requirements associated with each strategy of each task as follows
\begin{equation}
    \prescript{r}{}{\hat{y}_m} = \sum_{y_m^{(i)} \in \prescript{r}{}{C_m}} \frac{y_m^{(i)}}{\vert \prescript{r}{}{C_m} \vert}  
\end{equation}
where $\vert \prescript{r}{}{C_m} \vert$ denotes the number of demonstrations for Task $T_m$ that are identified as part of the $r^\text{th}$ cluster. Utilizing the average trait aggregation as an approximation of the task requirements associated with each strategy helps deal with potential noise in the demonstrations.

Our notion of heterogeneity in strategies does not point to the fact that different coalitions can carry out the task in a similar fashion (e.g., helicopters or fixed-wing aircraft can be used to aerially search an area). Rather, we reserve the term ``strategies" to denote fundamentally different, yet equivalent, approaches to carrying out the task (search the area aerially vs. from the ground). 

Revisiting our disaster response example from before, we could extract two distinct strategies from multiple expert coalitions assigned to the search task: one requiring large sensing radius and slow speed, and another requiring small sensing radius and high speed.

\subsection{Resource-aware coalition formation}

Given a new target team with the Species-Trait matrix $Q^{(j)} \notin \mathcal{D}$, we turn to the problem of optimizing its coalitions such that the inferred task requirements are satisfied. To this end, we formulate a constrained optimization problem capable of \textit{simultaneously} \textit{i)} choosing strategies for each of the $M$ tasks based on the resources available to the team, and \textit{ii)} optimizing robot assignment such that the requirements associated with the chosen strategies are met.

To represent the choice between multiple strategies, we introduce the \textit{strategy selectors} $z_m \in \{0,1\}^{P_m},\forall m$, one-hot encoded decision variables that identify the choice of strategy for each task. In addition, we introduce integer decision variables $x_m \in \mathbb{Z}^{S}_+, \forall m$ representing the assignment of robots to each task. 
We treat each task as a sub-problem and simultaneously optimize the overall assignment such that the chosen set of trait requirements are satisfied for all tasks.

To quantify trait satisfaction, we define an error measure called trait mismatch error as follows. If the $r^{\text{th}}$ strategy is chosen for Task $T_m$, the corresponding Trait Mismatch Error $\prescript{r}{}{e}_m$, between the aggregated traits and the trait requirements of the $r^{\text{th}}$ strategy is given by
\begin{equation}\label{eq:mismatch}
    \prescript{r}{}{e}_m = \Vert \prescript{r}{}{\hat{y}}_m - {Q^{(j)}}^T x_m \Vert_2^2
\end{equation}
where $x_m$ represents the assignment for the Task $T_m$.

Based on the definitions above, we define 
the net trait mismatch error for Task $T_m$ as
\begin{align}
    E_m & = z_m^T\ e_m \label{eq:cost}
\end{align}
where $e_m = [\prescript{1}{}{e_m},\cdots,\prescript{P_m}{}{e_m}]^T \in \mathbb{R}^{P_m}_+$ is a vector of trait mismatch errors between the aggregated traits and each of the strategies' task requirements.

Finally, we cast the \textit{resource-aware} optimization of robot assignments for the new team with $Q^{(j)} \notin \mathcal{D}$ as a constrained quadratic integer program:
\begin{align}
    \{x_m^{*(j)}, z_m^{*(j)}\}_{m=1}^{M} & = \arg \min_{x_m, z_m}\ \sum_m E_m \label{eq:opt} \\
    \mathrm{s.t.} & \sum_m x_m \leq N_a \label{eq:num_agent_consv} \\
    & z_m^T \cdot \bm{1} = 1,\     \forall m \label{eq:one-strat}\\
    & \hat{Y}_m z_m \leq {Q^{(j)}}^T x_m,\ \forall m \label{eq:desired-task-req}
\end{align}
where $N_a \in \mathbb{Z}_+^S$ represents the vector of total robots per species, $\bm{1}$ is a vector of ones, and $\hat{Y}_m = [\prescript{1}{}{\hat{y}_m},\cdots, \prescript{P_m}{}{\hat{y}_m}] \in \mathbb{R}_+^{U \times P_m}$ represents all the distinct trait requirements for Task $T_m$ extracted from the demonstrations.

Our approach is resource-aware since it optimizes both task allocation and strategy selection while taking into account the resources available to the target team as given by $Q^{(j)}$. As shown in (\ref{eq:num_agent_consv}), we explicitly constrain our approach to not recruit more robots than available in each species. The constraint in (\ref{eq:one-strat}) ensures that only one strategy is chosen for each task. Finally, the constraint in (\ref{eq:desired-task-req}) ensures that the assignment satisfies the desired trait requirements associated with the chosen strategies. However, a notable limitation of our approach is that the optimization problem above is not guaranteed to converge to the global optimum. 

In scenarios with under-resourced teams, the optimization problem defined in (\ref{eq:opt})-(\ref{eq:desired-task-req}) might be infeasible. In such cases, we relax the problem by removing the constraints in (\ref{eq:desired-task-req}) as described in Algorithm \ref{alg:exep}. This relaxation allows our approach to adaptively choose strategies and minimize the trait mismatch error for under-resourced teams.

Revisiting our disaster response example, our framework chooses between the two strategies for the search task (low speed and large sensing radius vs. high speed and small sensing radius) depending on the capabilities of the available team. Indeed, one of the two strategies is likely to be better suited than the other for a given target team. Further, the resource constraint in (\ref{eq:num_agent_consv}) helps our framework realize that if all ground vehicles are assigned to the search task, we will not be able to utilize them to remove debris.

\begin{algorithm}[t]
  \SetNoFillComment   
  \SetKwInOut{Input}{Input}
  \SetKwInOut{Output}{Output}
  \SetKwProg{try}{try}{:}{}
  \SetKwProg{catch}{except}{:}{end}
  \Input{$\{\prescript{r}{}{\hat{y}_m}\}, \forall r,m,$ and $ Q^{(j)} \notin \mathcal{D}$}
  \Output{$X^{*(j)}, Z^{*(j)}$} 
  
  Initialize int $x_m$ and $z_m$, $\forall m \in \{1,\cdots,M\}$\\

  \try{}{
    \tcc{Attempt to satisfy all trait requirements}
    Compute trait mismatch for each task from (\ref{eq:cost})\\
    Apply constraints (\ref{eq:one-strat}), (\ref{eq:num_agent_consv}), and (\ref{eq:desired-task-req})\\
    Optimize cost in (\ref{eq:opt}) subject to (\ref{eq:one-strat}), (\ref{eq:num_agent_consv}), and (\ref{eq:desired-task-req}) \\
  }
  
  \catch{}{
    \tcc{If under-resourced, relax minimum requirements}
    Compute trait mismatch for each task from (\ref{eq:cost})\\
    Apply constraints (\ref{eq:one-strat}) and (\ref{eq:num_agent_consv})\\
    Optimize cost in (\ref{eq:opt}), subject to (\ref{eq:one-strat}) and (\ref{eq:num_agent_consv}) \\
  }
  \Return $X^{*(j)}, Z^{*(j)}$
  \caption{Resource-aware coalition formation}
  \label{alg:exep}
\end{algorithm}

    


\section{EXPERIMENTAL EVALUATION}

We evaluate our approach in terms of its ability to satisfy trait requirements when forming coalitions for new teams using detailed numerical simulations, StarCraft II battles, and a multi-robot emergency-response scenario.
Specifically, we evaluate our framework's ability to generalize the inferred heterogeneous task allocation strategies to teams with different number of robots, and different types of species.

\subsection{Baselines}\label{subsec:conditions}

In all the experiments below, we compare the performance of our approach against the following baselines that take different factors into account:

\noindent\textit{1. No Heterogeneity} (\textcolor{NH}{\texttt{NH}}):
This baseline assumes the existence of a single set of task requirements.
For this baseline, we compute the trait requirements as an average across all demonstrations for each task:
\begin{equation}\label{eq:unimodal}
    y^*_{m_{uni}} = \frac{\sum_{j=1}^N y_m^{(j)}}{N}
\end{equation}
With only one strategy, task allocation is optimized by solving a constrained least squares problem. We included this baseline to validate the need to consider the heterogeneity in task requirements.
\vspace{4pt}

\noindent\textit{2. No Context} (\textcolor{NC}{\texttt{NC}}):
This baseline extracts multiple strategies from demonstrations like our approach. However, when optimizing for assignments, it picks strategies at random irrespective of the team's capabilities (i.e., the context) and solves a constrained least squares problem. We included this baseline to validate the need to consider the team's resources when selecting which strategy to pursue.
\vspace{4pt}

\noindent\textit{3. No Abstraction} (\textcolor{NA}{\texttt{NA}}):
This baseline does not cluster the demonstrations and solves the optimization problem in (\ref{eq:opt})-(\ref{eq:desired-task-req}) with every demonstration as a potential strategy. We included this baseline to validate the need to distill the demonstrations into a 
small number of strategies.
\vspace{4pt}

\noindent\textit{4. Random Allocation} (\textcolor{RA}{\texttt{RA}}):
This baseline entirely ignores the notion of task requirements and assigns a random subset of the available robots to each task without overlap. We included this baseline to validate the need for careful task allocation.

\noindent\textit{5. Behavioral Cloning} (\textcolor{BC}{\texttt{BC}}):
This baseline is an imitation learning algorithm \cite{osa2018algorithmic} which mimics expert policies directly. For coalition formation, we designed a behavioral cloning baseline that trains a neural network to approximate the mapping from the capabilities of the team $Q$ to the appropriate assignment $X$. We used the demonstrations as training data for the network.

We note that both \textcolor{NH}{\texttt{NH}} and \textcolor{NC}{\texttt{NC}} baselines resemble 
prior work in coalition formation (e.g., \cite{vig2006multi, duvallet2010imitation,cui2017learning,ravichandar2020strata}) as they rely on a single \textit{pre-specified} set of task requirements. The \textcolor{BC}{\texttt{BC}} baseline on the other hand is similar to the approach from \cite{duvallet2010imitation}, and more broadly represents a 
naive approach to learn coalition formation from expert demonstrations.

In contrast to all baselines, our structured approach embodies all three considerations (heterogeneity, context, and abstraction) and solves the optimization problem in (\ref{eq:opt})-(\ref{eq:desired-task-req}) after inferring the strategies from the demonstrations.


\subsection{Metrics}\label{subsec:metrics}

We evaluate each allocation using the following metrics:

\noindent\textit{1. Minimum trait mismatch} is defined as the smallest difference between the achieved trait aggregation and any of the potential task requirements, if the achieved trait aggregation is lesser than all the potential task requirements, otherwise it is 0. In other words, minimum trait mismatch does not penalize over-provisioning and is given by,
\begin{equation}\label{eq:mintraiterr}
    E_{\text{min}} (\hat{x}_m) = 
    \begin{cases}
    \frac{\min_r (\Vert \prescript{r}{}{\hat{y}_m} - Q^{T} \hat{x}_m \Vert_2)}{\Vert \prescript{r}{}{\hat{y}_m} \Vert_2} & \text{if $\prescript{r}{}{\hat{y}_m} > Q^{T} \hat{x}_m$} \\
    0 & \text{Otherwise}
    \end{cases}
\end{equation}
where $\hat{x}_m$ denotes the assignment for Task $T_m$ computed by an approach, and $Q$ represents the species-trait matrix of the target team. 
\vspace{4pt}

\noindent\textit{2. Exact trait mismatch} is defined as the smallest difference between the achieved trait aggregation and any of the potential task requirements, given by
\begin{equation}\label{eq:extraiterr}
    E_{\text{exact}}(\hat{x}_m) = 
    \frac{\min_r (\Vert \prescript{r}{}{\hat{y}_m} - Q^T \hat{x}_m \Vert_2)}{\Vert \prescript{r}{}{\hat{y}_m} \Vert_2}
\end{equation}
We compute the smallest difference since all strategies are equivalent and it is sufficient to satisfy the requirements associated with any one of them. 
\vspace{4pt}

\noindent\textit{3. Robot utilization:} In addition to trait mismatch errors, we also define a metric to account for the relative number of robots recruited by each algorithm as 
\begin{equation}\label{eq:util}
    RU(\hat{x}_m) = \frac{\hat{x}_m^T \cdot \bm{1}}{\mathcal{N}_a \cdot \bm{1}}
\end{equation}

\noindent\textit{4. Success Rate:} Finally, we measure the team's performance in experiments  from Sections \ref{subsec:StarCraft-II} and \ref{subsec:Robotarium}, based on their success or failure in executing the tasks: 

\begin{equation}\label{eq:taskperf}
    SR(\hat{x}_m) = \frac{N_{sucess}(\hat{x}_m)}{N_{runs}} \times 100
\end{equation}
where $N_{success}(\hat{x}_m)$ denotes the number of times the tasks were successfully executed out of $N_{runs}$ attempts. We also report the computational costs associated with each approach in Appendix \ref{app:comp_cost}.

Combined, these metrics measure the ability of an approach to effectively optimize the allocation such that the corresponding task requirements are met while recruiting the least number of robots. We statistically analyze all our results using the Kruskal-Wallis test, followed by the Dunn test for post-hoc pairwise comparisons and FDR adjustment. 

\subsection{Numerical analyses}\label{subsec:Numerical_analyses}

We first evaluate all the approaches from Section \ref{subsec:conditions} using the metrics from Section \ref{subsec:metrics} when choosing the coalitions for simulated target teams that were not encountered in the demonstrations. These detailed numerical simulations help analyze the performance of each algorithm across a wide variety of problems by altering aspects such as team size, robot capabilities, and task strategies. See Appendix \ref{app:numerical} for additional experiments.

\subsubsection*{Design}
We consider simulated task allocation problems, each with four species ($S=4$), three traits ($U=3$), three tasks ($M=3$), three strategies per task ($P_m = 3, \forall m$), and the number of robots per species uniformly randomly sampled between $6$ and $33$. We generate a set of $240$ demonstrations ($\mathcal{D}=\{X^{(i)}, Q^{(i)}\}_{i=1}^{240}$) and test our approach and the baselines on $60$ target teams ($\{Q^{(j)}\}_{j=241}^{300}$). We generate teams and strategies such that our data contains a mix of under-, sufficiently-, and over-resourced teams. We set $1800$ seconds as an upper bound on the optimization time for all approaches.

\subsubsection*{Results and discussion}
\begin{figure}
    \centering
    \includegraphics[width=\columnwidth]{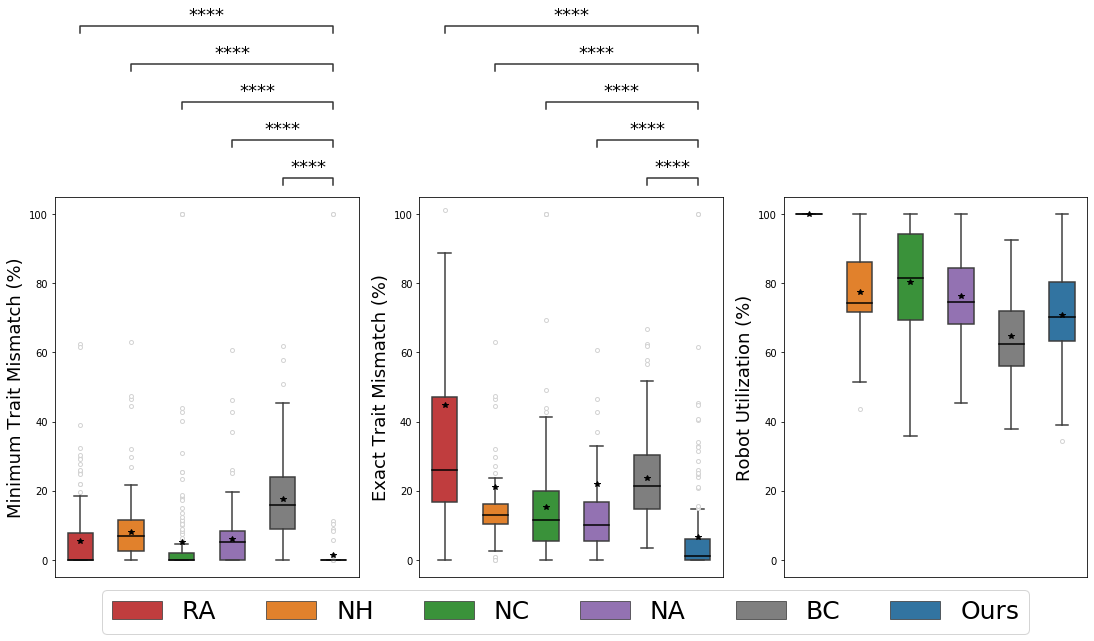}
    \caption{The figure shows subplots of measures of minimum trait mismatch, exact trait mismatch, and robot utilization (from left to right) observed in the numerical analyses. As shown, our approach achieves the lowest minimum and exact trait mismatch percentage error computed across $3$ tasks over $60$ test teams. We performed the Kruskal-Wallis test, followed by the Dunn test for post-hoc pairwise comparisons and FDR adjustment and found that all comparisons are statistically significant ($p < 1e-5$).}
    \label{fig:exp1}
\end{figure}
First, our approach was able to extract three distinct strategies for each task from the demonstration set $\mathcal{D}$. Given the inferred strategies, we evaluated each algorithm's ability to optimize the coalitions of the target team in terms of the metrics in Section \ref{subsec:metrics}. 
As shown in Fig. \ref{fig:exp1}, the \textcolor{RA}{\texttt{RA}} baseline consistently recruits all the available robots and yet results in the highest $E_{\text{exact}}$ and a non-trivial $E_{\text{min}}$. In contrast, our approach (\textcolor{ours}{\texttt{ours}}) and all other baselines (\textcolor{NH}{\texttt{NH}}, \textcolor{NC}{\texttt{NC}}, \textcolor{NA}{\texttt{NA}}, and \textcolor{BC}{\texttt{BC}}) utilize fewer proportion of robots and result in comparable or lower errors. These observations demonstrate the need for careful task allocation.

We find that our approach (\textcolor{ours}{\texttt{ours}}) outperforms all the baselines (\textcolor{NH}{\texttt{NH}}, \textcolor{NC}{\texttt{NC}}, \textcolor{NA}{\texttt{NA}}, and \textcolor{BC}{\texttt{BC}}) in terms of both minimum and exact trait mismatch errors, and that the improvements are statistically significant ($p < 1e-5$). 
Further, our approach is able to do so while utilizing a comparable number of robots. This observation indicates that our approach utilizes the recruited robots more effectively than all the baselines. We also observe that ignoring heterogeneity (\textcolor{NH}{\texttt{NH}}) leads to poorer trait satisfaction as indicated by higher $E_{\text{min}}$, and failing to consider the team's resources (\textcolor{NC}{\texttt{NC}}) results in over-provisioning as indicated by higher $E_{\text{exact}}$. This observation points to the deficiencies of existing coalition formation approaches that either ignore heterogeneous strategies or the context of available resources. 

We also find that ignoring abstraction (\textcolor{NA}{\texttt{NA}}) incurred a considerably larger computational cost ($1067.67\pm{441.47}$s) than our approach ($11.12\pm{26.94}$s)\footnote{See Appendix D 
for more details}. This increased computational burden of \textcolor{NA}{\texttt{NA}} is a natural result of considering all data points in $\mathcal{D}$ as a unique strategy. In contrast, our approach (\textcolor{ours}{\texttt{ours}}) reasons over a 
smaller number of distinct strategies that are distilled from the demonstrations.

Finally, behavioral cloning (\textcolor{BC}{\texttt{BC}}) has the highest minimum trait mismatch performing (worse than randomly allocating agents), and the second highest exact trait mismatch. 
This is consistent with robot utilization metric showing that \textcolor{BC}{\texttt{BC}} allocated the least proportion of robots. These findings illustrate that unstructured supervised learning using available demonstrations is an ineffective technique to learn coalition formation. 

\subsection{Evaluation on StarCraft II}\label{subsec:StarCraft-II}

In the next set of experiments, we evaluate all the approaches from Section \ref{subsec:conditions} on StarCraft II battles using the metrics from Section \ref{subsec:metrics}. The agents in the game belong to different species, each with a particular sets of capabilities as listed in Table \ref{tab:species-trait} in Appendix \ref{app:starcraft}. 
We handle trajectory-level control of the agents by using in-built AI bots from StarCraft II editor to ensure that, differences in their individual actions do not interfere with our assessment of the coalitions' performance.
The battles emulate tasks that require careful allocation using combinations of the available species. The battles also aid in the evaluation of our approach against the baselines in terms of task-level performance. To compare the performance of all approaches, we compute the percentage success rate $SR$ as defined in (\ref{eq:taskperf}). 

\begin{figure}
    \centering
    \includegraphics[width=\columnwidth]{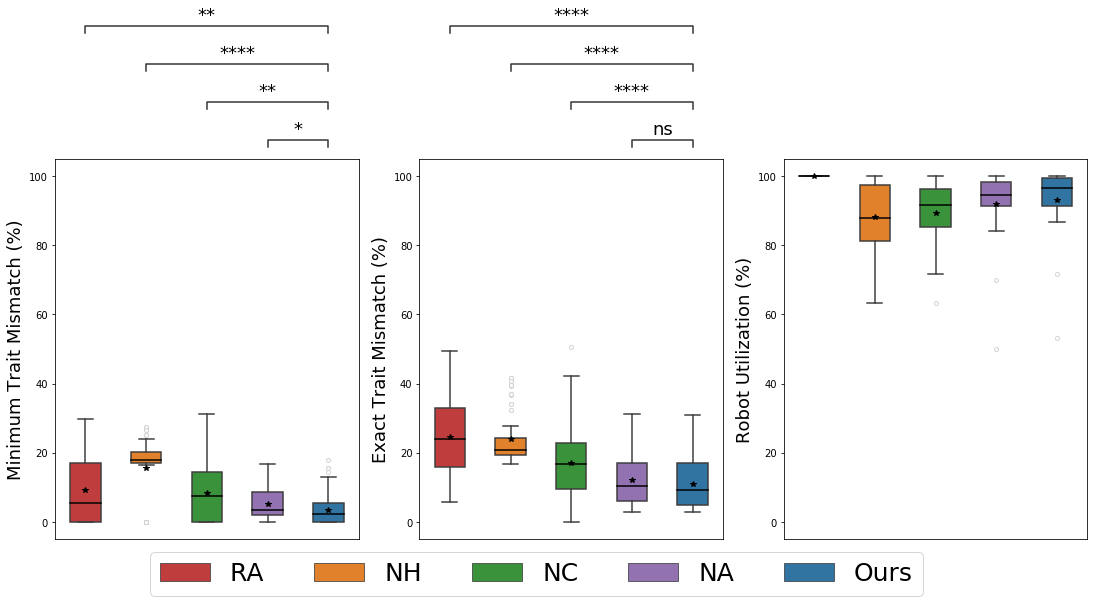}
    \caption{Measures of trait satisfaction and robot utilization on StarCraft battles (lower is better).}
    \label{fig:exp2err}
\end{figure}
\begin{figure}
    \centering
    \includegraphics[width=\columnwidth]{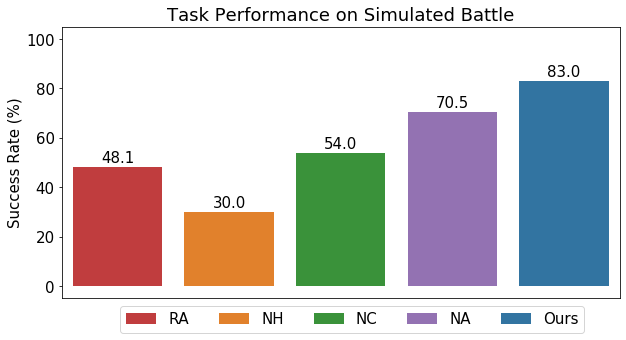}
    \caption{Success rate of our approach and that of the baselines on StarCraft battles (higher is better).}
    \label{fig:winrate}
\end{figure}

\subsubsection*{Design}
We design simulated battles based on the 2s3z map from the StarCraft II Multi-Agent Challenges (SMAC)~\cite{samvelyan2019starcraft}. Specifically, we define four concurrent tasks ($M=4$) on the battle map, each involving a battle with 10 Stalkers and 15 Zealots.

We generate 9 demonstrations ($\mathcal{D}=\{X^{(i)}, Q^{(i)}\}_{i=1}^{9}$) representing three strategies ($P_m = 3$) (as shown in Table \ref{tab:strat1}), each involving four different species ($S=4$), for the battle against 10 Stalkers and 15 Zealots. We then test each approach on a set of $10$ target teams ($\{Q^{(j)}\}_{j=10}^{19}$), each consisting a subset of five total available species. Further details of the experimental design can be found in 
Appendix \ref{app:starcraft}.

Note that \textcolor{BC}{\texttt{BC}} cannot generalize to any new species as the trained network cannot handle inputs of varying size. Further, \textcolor{BC}{\texttt{BC}} will likely generate invalid allocations that violate important resource constraints, such as the number of agents available in each species. As such, we do not include \textcolor{BC}{\texttt{BC}} in these experiments. Further details of the experimental design can be found in 
Appendix \ref{app:starcraft}.

\subsubsection*{Results and discussion}
Our approach identified three strategies using Agglomerative Clustering from the 9 demonstrations. Strategy 2 has the highest Att. (A) and DPS (A) which measure the ability of the coalition to attack the enemy from the air while Strategy 3 has no aerial attack power but makes up for it with the highest armor and health capabilities. Strategy 1 has moderate attack and defensive capabilities when compared to 2 and 3. On evaluating the coalitions optimized by the approaches from the inferred strategies, 
we note that the trends in terms of trait mismatch and robot utilization are similar to those from the numerical experiments. 

As shown in Fig. \ref{fig:exp2err}, the \textcolor{RA}{\texttt{RA}} baseline results in the highest $E_{\text{exact}}$ and a non-trivial $E_{\text{min}}$, even when utilizing all the available robots. In contrast, our approach (\textcolor{ours}{\texttt{ours}}) and all other baselines (\textcolor{NH}{\texttt{NH}}, \textcolor{NC}{\texttt{NC}}, and \textcolor{NA}{\texttt{NA}}) utilize fewer proportion of robots and result in comparable or lower errors. These observations demonstrate the need for careful task allocation.

We find that our approach (\textcolor{ours}{\texttt{ours}}) outperforms baselines (\textcolor{NH}{\texttt{NH}} and \textcolor{NC}{\texttt{NC}}) in terms of both minimum and exact trait mismatch errors, and that the improvements are statistically significant ($p < 0.01$). Further, our approach is able to do so while utilizing a comparable number of robots. This observation indicates that our approach utilizes the recruited robots more effectively. 

Finally, ignoring abstraction (\textcolor{NA}{\texttt{NA}}) 
incurred a considerably larger computational cost ($13.22\pm{7.82}$s) than our approach ($1.39\pm{1.17}$s)\footnote{See Appendix \ref{app:comp_cost} for more details}. These findings suggest that accounting for heterogeneity, context, and appropriate abstraction results in higher computational efficiency and better satisfaction of task requirements.

On comparing the success rates reported in Fig. \ref{fig:winrate}, the \textcolor{NH}{\texttt{NH}} baseline performs worse than all other baselines including \textcolor{RA}{\texttt{RA}}. This observation points to the fact that ignoring heterogeneity and relying on statistical averages of multi-modal distributions could lead to adverse effects. Further, a potential explanation for the performance of \textcolor{RA}{\texttt{RA}} baseline could be the existence of additional strategies that were not represented in the demonstrations. Note that, even in such circumstances, our approach is able to select one of the inferred strategies that is best suited for the team.

Generally, approaches with lower $E_\text{min}$ resulted in higher success rates. Our approach results in the highest success rate across all tasks and only resulted in two failures, both due to under-resourced teams. While the approach that ignores abstraction (\textcolor{NA}{\texttt{NA}}) had the second highest win rate, it is considerably more computationally expensive. In summary, our approach (\textcolor{ours}{\texttt{ours}}) outperforms all the baselines in terms of both trait satisfaction and task performance without incurring significant computational burden.

\subsection{Evaluation on Robotarium}\label{subsec:Robotarium}

In our final set of experiments, we design a multi-robot emergency-response scenario on the Robotarium, a publicly-accessible testbed~\cite{wilson2020robotarium}. 

\subsubsection*{Design} 
We designed the following three tasks for the emergency-response mission:

\begin{itemize}
    \item \textit{Move debris}: Two to five robots form a coalition and move a piece of debris weighing 50 pounds to a goal location. 
    The capabilities needed for the task are payload capacity and mobility. 
    
    \item \textit{Search an urban environment}: Two to five robots search an urban environment to look for survivors. The capabilities needed for the task are coverage area and mobility. 
    
    \item \textit{Retrieve objects from a narrow passage}: Robots are required to navigate and retrieve objects from a narrow passage. The capabilities needed for the task are miniature size, coverage area, and mobility. 
\end{itemize}
The strategies for all three tasks involve the use of either ground robots or aerial robots as shown in Table \ref{tab:robstrat} of Appendix \ref{app:robotarium}.

The test team consists of 5 ground robots, 4 aerial robots, 3 miniature ground robots and 1 miniature aerial robot. We evaluate all the approaches on the test team by computing 100 possible coalition-task pairs for every task. As mentioned in Section \ref{subsec:StarCraft-II}, we do not include \textcolor{BC}{\texttt{BC}} as a baseline in these experiments due to the learned networks inability to i) handle inputs of a difference size and ii) obey resource constraints.

\subsubsection*{Results and discussion}

\begin{figure}
    \centering
    \includegraphics[width=\columnwidth]{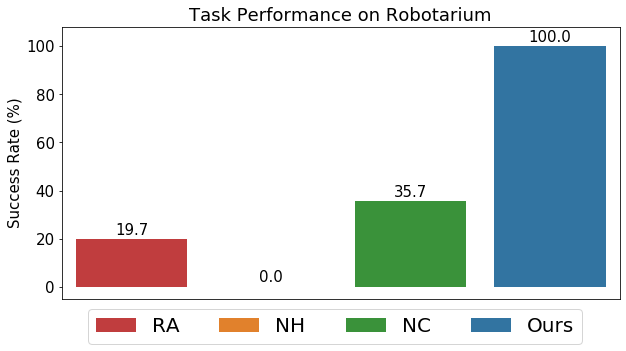}
    \caption{Success rate of our approach and that of the baselines on robotarium tasks (higher is better).}
    \label{fig:exp3win}
\end{figure}



On comparing the task performance as shown in Fig. \ref{fig:exp3win}, our approach (\textcolor{ours}{\texttt{ours}}) consistently assigns teams that satisfy the trait requirements across all tasks, achieving a 100\% success rate. The \textcolor{RA}{\texttt{RA}} baseline fails to assign meaningful coalitions in most cases and results in low success rate (19.7\%) indicating that the tasks require careful allocation. Notably, all the coalitions assigned by the \textcolor{NH}{\texttt{NH}} baseline fail because computing the statistical average of all demonstrations does not result in meaningful trait requirements. Finally, the \textcolor{NC}{\texttt{NC}} baseline performs better than both \textcolor{NH}{\texttt{NH}} and \textcolor{RA}{\texttt{RA}} baselines, but results in a considerably lower success rate (35.7\%) compared to our approach, indicating that context-dependent selection of strategies is necessary. 

Taken together, the three experiments validate the need for the different components of our framework (\textcolor{ours}{\texttt{ours}}) and demonstrate that our approach outperforms all baselines (\textcolor{NH}{\texttt{NH}}, \textcolor{NC}{\texttt{NC}}, \textcolor{NA}{\texttt{NA}}, \textcolor{BC}{\texttt{BC}}, and \textcolor{RA}{\texttt{RA}}) across all experimental conditions.

\section{CONCLUSIONS}

We proposed an approach to learn coalition formation strategies when task requirements are not explicitly known. Our framework was capable of distilling expert demonstrations into \textit{heterogeneous strategies} that capture different, equally-valid trait requirements. Further, our framework was effective in generalizing the inferred strategies to new target teams by considering the capabilities of the specific target team (i.e., context) without requiring additional training. Detailed experiments using numerical simulations, StarCraft II battles, and a multi-robot emergency-response mission reveal that our approach outperforms existing approaches to coalition formation that ignore i) the heterogeneity in strategies, ii) the context of available resources, or iii) appropriate abstraction to extract a 
small number of strategies. Future work will address key limitations of our framework, which include reliance on high-quality expert demonstrations, and inability to generalize to new tasks.



\begin{appendices}

\section*{APPENDICES}

\renewcommand{\thesectiondis}[2]{\Roman{section}:}

\section{Additional numerical analysis}\label{app:numerical}

\begin{figure}
    \centering
    \includegraphics[width=\columnwidth]{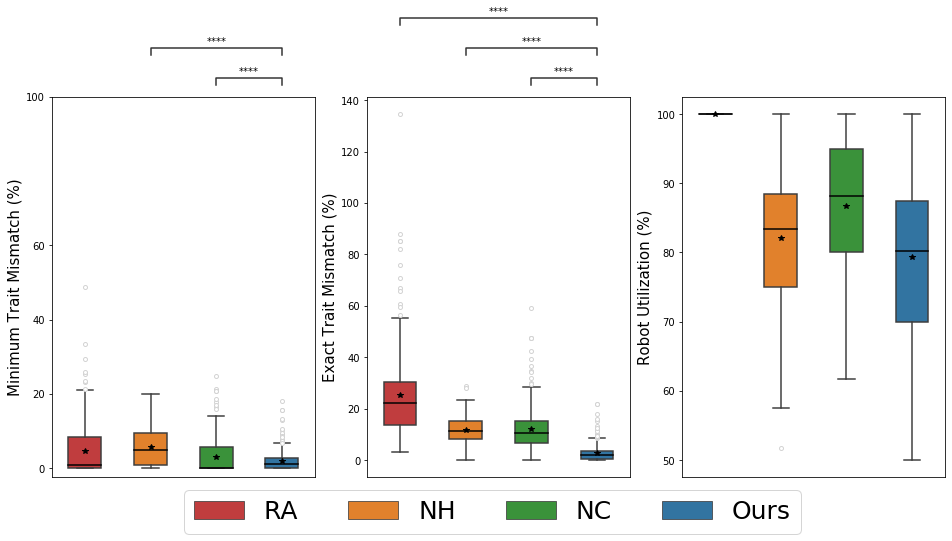}
    \caption{Performance, as measured by trait satisfaction and robot utilization percentages, of our approach and that of the baselines on 20 datasets (lower is better).}
    \label{fig:exp1numsim}
\end{figure}

In addition to the numerical experiments reported in Section \ref{subsec:Numerical_analyses}, we conducted additional numerical experiments by varying the number of species and the number of tasks. We considered 20 such datasets by varying the number of available species from $\{2, 4, 6, 8, 10\}$ and the number of concurrent tasks from $\{2, 3, 4, 5\}$. We report that the minimum and exact trait mismatch errors follow similar trends as seen in Fig. \ref{fig:exp1numsim}. The baselines considered for this experiment are namely \textcolor{RA}{\texttt{RA}}, \textcolor{NH}{\texttt{NH}} and \textcolor{NC}{\texttt{NC}}. The trait requirements for each task is generated by sampling from a Gaussian mixture model with 3 clusters.   

\section{Details of StarCraft II experiments}\label{app:starcraft}

For the design of StarCraft II experiments, we use 4 species from the game, namely Zealots, Stalkers, Marines and Marauders and simulate demonstrations by specifying different combinations of ally teams that lead to wins against the enemy team. The traits of the different species are listed in Table \ref{tab:species-trait}. We test our approaches with an additional species not seen in the demonstrations namely Roach. 

\begin{table}[!h]
\caption{Traits of Species for StarCraft II Experiments}\label{tab:species-trait}
\centering
\begin{tabularx}{\columnwidth}{|X|X|X|X|X|X|}
 \hline
 Traits&Zealot&Stalker&Marine&Marauder&Roach\\
 \hline
 Armor   & 1 & 1 & 0 & 1 & 1\\
 Health  & 100 & 80 & 45 & 125 & 145\\
 Shield   & 50 & 80 & 0 & 0 & 16\\
 Att. (G) & 8 & 13 & 6 & 5 & 16\\
 Att. (A) & 0 & 13 & 6 & 0 & 0\\
 DPS (G) & 18.6 & 9.7 & 9.8 & 9.3 & 11.2\\
 DPS (A) & 0 & 9.7 & 9.8 & 0 & 0\\
 Cooldown & 0.86 & 1.34 & 0.61 & 1.07 & 1.43\\
 Speed & 3.15 & 4.13 & 3.15 & 3.15 & 3.15\\
 Range & 0 & 6 & 5 & 6 & 4\\
 Sight & 9 & 10 & 9 & 10 & 9\\
 \hline
\end{tabularx}
\end{table}

For the design of traits that are non-cumulative such as Cooldown, we compute the inverse to make it cumulative. The Speed trait threshold is set at 4, which implies that species that have movement speed greater than 4 assume the binary value 1 and it is 0 otherwise.

\begin{table}[!h]
\caption{Inferred Strategies for Battle (10s15z)}\label{tab:strat1}
\begin{tabularx}{\columnwidth}{|X|X|X|X|}
 \hline
 Traits&Strategy 1&Strategy 2&Strategy 3\\
 \hline
 Armor   & 28.67 & 0 & 30\\
 Health  & 2566.67 & 2010 & 3225\\
 Shield   & 1883.33 & 0 & 1050\\
 Att. (G) & 304.33 & 268 & 213\\
 Att. (A) & 195 & 268 & 0\\
 DPS (G) & 399.7 & 223.33 & 474.3\\
 DPS (A) & 145.5 & 437.73 & 0\\
 Cooldown & 27.09 & 73.22 & 32.83\\
 Speed & 15 & 0 & 0\\
 Range & 90 & 223.33 & 54\\
 Sight & 273 & 402 & 279\\
 \hline
\end{tabularx}
\end{table}

\section{Details of Robotarium experiments}\label{app:robotarium}

The strategies for all tasks are mentioned in Table \ref{tab:robstrat}. For task 1, the debris can be moved by either a coalition of ground robots or aerial robots with collective payload capacity meeting the trait requirements. Hence, we have ground robots and aerial robots encoded as separate traits. The total area to be covered to search an urban environment in task 2 is 8 $m^2$ and can be achieved using ground or aerial robots. In task 3, either miniature ground robots with higher coverage area or miniature aerial robots with smaller coverage area can be used to retrieve objects.

\begin{table}[!h]
\caption{Inferred Strategies for Tasks on Robotarium}\label{tab:robstrat}
\begin{tabularx}{\columnwidth}{|X|X|X|X|X|X|}
 \hline
 &Coverage Area ($m^2$)&Ground Robots&Payload Capacity ($lb$)&Miniature Quality&Aerial\ \ \ \ Robots\\
 \hline
 Task 1 & 0 & 0 & 50 & 0 & 5\\
        & 0 & 5 & 50 & 0 & 0\\
 \hline
 Task 2 & 8 & 4 & 0 & 0 & 0\\
        & 8 & 0 & 0 & 0 & 4\\
 \hline
 Task 3 & 6 & 3 & 0 & 3 & 0\\
        & 6 & 0 & 0 & 6 & 6\\
 \hline
\end{tabularx}
\end{table}

\section{Computational costs}\label{app:comp_cost}

Both the numerical simulation and StarCraft II experiments were performed on an Intel i7(4GHz) CPU with 16GB memory. The numerical simulations were tested on 60 test teams (problem instances) to be assigned to three tasks. On StarCraft II, the experiment was performed on 10 target teams to be assigned to four simultaneous battles. In Table \ref{tab:comp-time}, we report the average and standard deviation of the run times in seconds for the different approaches (for solving a single problem instance). In Table \ref{tab:mem}, we report the CPU memory space required in mebibyte (MiB) for the different approaches obtained from the memory profiler.

\begin{table}[!h]
\caption{Computation Time for the different approaches}\label{tab:comp-time}
\centering
\begin{tabularx}{\columnwidth}{|c|c|c|X|X|}
 \hline
 &NH (s) &NC (s) &NA (s) &Ours (s)\\
 \hline
 Num.  & 8.68$\pm{34.98}$ & 33.53$\pm{135.18}$ & 1067.7$\pm{441.5}$ & 11.12$\pm{26.94}$ \\
 SC II  & 1.56$\pm{3.31}$ & 0.21$\pm{0.22}$ & 13.22$\pm{7.82}$ & 1.39$\pm{1.17}$ \\
 \hline
\end{tabularx}
\end{table}
\begin{table}[!h]
\caption{Memory required for the different approaches}\label{tab:mem}
\centering
\begin{tabularx}{\columnwidth}{|c|X|X|X|X|}
 \hline
 &NH (MiB) &NC (MiB) &NA (MiB) &Ours (MiB)\\
 \hline
 Num.  & 169.766 & 172.195 & 175.977 & 164.980 \\
 SC II  & 196.012 & 191.504 & 203.508 & 192.254 \\
 \hline
\end{tabularx}
\end{table}

\end{appendices}





\bibliographystyle{ieeetr}
\bibliography{mrs}

\end{document}